\documentclass{article}
\usepackage{spconf,amsmath,graphicx,hyperref}

\usepackage{graphicx}

\usepackage{cite}
\usepackage{amsmath,amssymb,amsfonts}
\usepackage{algorithmic}
\usepackage{textcomp}

\usepackage{booktabs}       
\usepackage{multirow}
\usepackage{float}
\usepackage[dvipsnames]{xcolor}
\usepackage{colortbl}
\usepackage{bm}
\usepackage{arydshln}
\usepackage{makecell}
\usepackage{siunitx}

\title{DiTSE: High-Fidelity Generative Speech Enhancement via \\ Latent Diffusion Transformers}
\name{\begin{tabular}{c}Heitor R. Guimar\~aes$^{1, \dagger}$, Jiaqi Su$^{2}$, Rithesh Kumar$^{2}$, Tiago H. Falk$^{1}$, Zeyu Jin$^{2}$\end{tabular}\thanks{$^{\dagger}$ Work done during an internship at Adobe Research.}}

\address{%
    $^{1}$ INRS-EMT, Université du Québec, Montréal, Canada\\%
    $^{2}$ Adobe Research, San Francisco, California, United States
}

\begin{document}
\ninept
\maketitle
\begin{abstract}
Real-world speech recordings suffer from degradations such as background noise and reverberation. Speech enhancement aims to mitigate these issues by generating clean high-fidelity signals. While recent generative approaches for speech enhancement have shown promising results, they still face two major challenges: (1) content hallucination, where plausible phonemes generated differ from the original utterance; and (2) inconsistency, failing to preserve speaker's identity and paralinguistic features from the input speech. In this work, we introduce \textbf{DiTSE} (Diffusion Transformer for Speech Enhancement), which addresses quality issues of degraded speech in full bandwidth. Our approach employs a latent diffusion transformer model together with robust conditioning features, effectively addressing these challenges while remaining computationally efficient. Experimental results from both subjective and objective evaluations demonstrate that DiTSE achieves state-of-the-art audio quality that, for the first time, matches real studio-quality audio from the DAPS dataset. Furthermore, DiTSE significantly improves the preservation of speaker identity and content fidelity, reducing hallucinations across datasets compared to state-of-the-art enhancers. Audio samples are available at: \textcolor{blue}{\url{http://hguimaraes.me/DiTSE}}
\end{abstract}


\section{Introduction}
\label{sec:intro}
Real-world recorded speech is often degraded by acoustic conditions such as noise, reverberation, and equalization issues. Speech enhancement (SE) methods aim to improve such signals' perceptual quality and intelligibility, with useful applications to audio content creation and telecommunications. Recent advances in deep learning have established new performance standards for speech enhancement. While existing methods can alleviate some quality issues, achieving clean high-fidelity audio remains a challenge~\cite{o2024speech}.

The conventional enhancement methods based on deep neural networks learn a deterministic many-to-one mapping via a regressive objective, which maps input signals that have been variously degraded from a clean signal to the corresponding ground-truth clean signal. However, the problem becomes ill-posed in challenging scenarios where degradations, such as reverberation, create ambiguity in determining the target phase of the clean signal. Therefore, the conventional methods often experience a regression-to-the-mean behavior, resulting in artifacts and distortion in their enhanced results.

Generative methods directly model the distribution of clean speech, avoiding mode collapse and enabling effective utilization of speech priors. They aim to generate a plausible version of many possible high-fidelity clean signals. Recently, discrete token generation methods~\cite{wang2023selm, xue2023low} have shown improvement in audio clarity and cleanness. Particularly, Genhancer~\cite{genhancer} incorporates speech features to improve speaker consistency and content accuracy. However, the token space often misses nuances in spoken content, and costly inference sampling limits real-world usability.

On the other hand, diffusion-based methods operate in a continuous space (i.e., sample or latent representation), stemming from the common concept of non-equilibrium thermodynamics of learning to reverse noise back to the data distribution iteratively. 
Particularly, SGMSE+~\cite{richter2023speech} and CDiffuSE~\cite{lu2022conditional} interpolate noisy speech with Gaussian noise via the forward diffusion process, aiming to reduce both the Gaussian noise and environmental degradations during the reverse process.  Alternatively, UNIVERSE~\cite{serra2022universal} and UNIVERSE++~\cite{universepp} model speech generation conditioning on the high-level features extracted from the input audio.
Moreover, StoRM~\cite{lemercier2023storm} leverages a diffusion process to refine the outputs from other enhancement systems via regeneration, combining benefits from both conventional and generative approaches. To enhance content fidelity and speaker similarity, several approaches inject speech features as conditioning signals within the diffusion process, as demonstrated by uSEE~\cite{10447017}. 
In addition, latent diffusion models have emerged as promising alternatives, offering improved efficiency and acoustic quality~\cite{dhyani2025high, kumar-etal-2025-prose}. Among these, a concomitantly released work on speech restoration, Hi-ResLDM~\cite{dhyani2025high}, is most closely related to our work. It adopts a two-stage architecture: first, a pretrained and frozen enhancement model recovers the degraded version of the signal; then, a latent diffusion model further refines it to improve perceptual quality.

In this work, we propose DiTSE (\textbf{Di}ffusion \textbf{T}ransformer for \textbf{S}peech \textbf{E}nhancement), a high-fidelity, full-bandwidth model that improves speaker preservation and reduces content hallucination over prior methods, while achieving state-of-the-art audio quality that, for the first time, matches real studio quality from DAPS~\cite{6981922}. DiTSE performs diffusion in the latent space of a variational autoencoder, along with a robust conditioning mechanism on speech features for content and speaker guidance. By leveraging Latent Diffusion Transformers, our method demonstrates strong performance and scalability. We include comprehensive objective and subjective evaluations on audio quality, content accuracy, and speaker consistency.

\begin{figure*}[t]
        \centering
        \includegraphics[width=\linewidth]{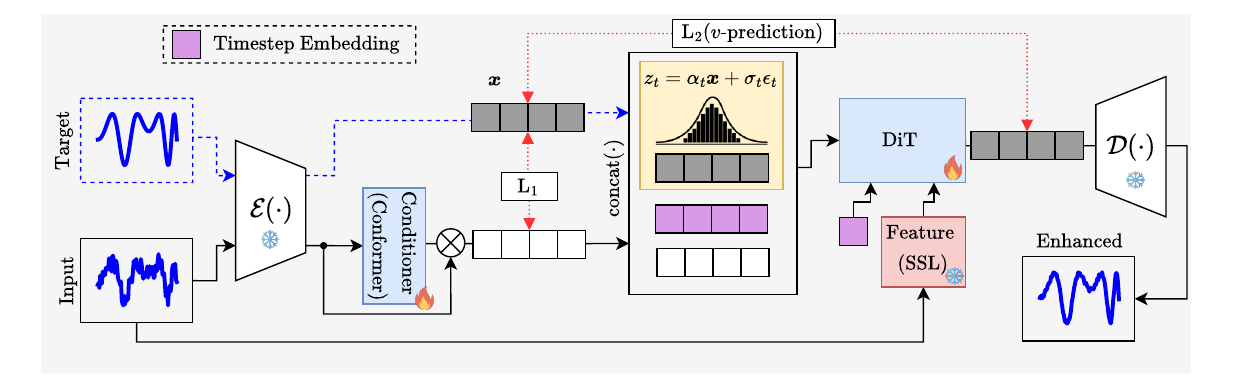}
        \vspace{-5mm}
        \caption{Diagram of DiTSE. The degraded input is encoded by the VAE and SSL models. The VAE latent undergoes pre-denoising to approximate clean latents, which, along with Gaussian noise and timestep embeddings, are fed into the DiT, conditioned on SSL embeddings via cross-attention. Lastly, the output is then decoded to reconstruct the waveform. Blue-dashed arrows denote training-only components; red-dotted arrows mark loss applications; black-solid arrows indicate paths used in both training and inference.}
        \label{fig:ditse_architecture}
\end{figure*}

\section{Proposed Method}
\label{sec:pagestyle}

DiTSE comprises three main components: (1) a variational autoencoder; (2) a diffusion transformer; and (3) a conditioner network that guides generation using both pre-trained and learned speech features. Additionally (and optional), a post-hoc refinement step improves audio quality via discrete token generation.

\subsection{Preliminaries on Diffusion Models}

Diffusion-based models have drawn attention due to their generation capabilities~\cite{evans2024stable}. Diffusion models employ a forward process that adds Gaussian noise to clean input data, which can be expressed in marginal $q_(\bm{z}_t | \bm{x}) = \mathcal{N}(\bm{z}_t | \alpha_t\bm{x}, \sigma^2_t\bm{I})$, where $\alpha_t, \sigma_t \in (0, 1)$ are hyperparameters from the noise scheduler~\cite{hoogeboom2023simple} at the timestep $t \in [0, 1]$. The objective is to learn to reverse this process via a neural network. However, instead of directly trying to predict the data reconstruction $\hat{\bm{x}}$, it has been empirically found that other objectives are more stable to produce high-quality outputs. In this work, we use the \textit{v-prediction} reparametrization~\cite{salimans2022progressive}, where the model predicts the velocity vector, $\bm{v}_t = \alpha_t \bm{\hat{\epsilon}}_t - \sigma_t \hat{\bm{x}}$, where $\bm{\hat{\epsilon}}_t \sim \mathcal{N}(0, 1)$. For the interested reader in the formulation, we refer to~\cite{hoogeboom2023simple}.

\subsection{Variational Autoencoder}
We train a variational autoencoder (VAE) on publicly available \SI{48}{kHz} audio data (speech, music, and sound events) to produce 64-dimensional latent representations at a \SI{40}{Hz} frame rate. The VAE follows the DAC~\cite{NEURIPS2023_58d0e78c} architecture, replacing its quantization layers with a variational bottleneck. To ensure high-fidelity reconstruction, it adopts DAC's GAN framework, including multi-period waveform and multi-scale multi-band spectrogram discriminators. The resulting latent space is both compressed and regularized, enabling efficient generation and high-quality audio synthesis.

\subsection{Latent Diffusion Transformer (DiT) Network}
While U-Net architectures commonly serve as backbones for diffusion-based speech enhancement methods~\cite{welker2022speech, richter2023speech}, recent advancements in audio generation~\cite{lovelace24_interspeech} have shown that Transformer-based backbones offer significant advantages in scalability and robustness. Herein, we adopt Diffusion Transformer (DiT)~\cite{peebles2023scalable} on the latent representations from the VAE, leveraging established practices and strategies from the vision domain for training and inference. The model conditions on the timestep embedding of the current diffusion step via adaptive layer normalization, following common practices. Other conditioning information (discussed below) is provided to the network either via cross-attention or via concatenation with the diffusion latent along the feature dimension.
During training, we apply prefix prompting with a 50\% probability, replacing up to the first 50\% of diffusion latent frames with the corresponding ground-truth clean latent. During enhancement of long audio, we provide a previously enhanced audio segment as a prompt to ensure consistency of generation across windows of windowed generation.

\subsection{Robust Conditioning Features}\label{subsec:cond}
The conditional speech generation should recover clean audio that faithfully preserves the original spoken content and paralinguistic features of the input audio. We take inspiration from the prior work Genhancer~\cite{genhancer} where the generative model conditions on two kinds of speech features: self-supervised learned (SSL) features extracted from the input audio, and a learned representation of the input audio.

\vspace{2mm}
\noindent\textbf{Robust SSL features}.
Leveraging large-scale pre-training on speech data, SSL features provide strong semantic guidance for spoken content, thus reducing content hallucinations in enhancement. This work experiments with WavLM~\cite{9814838} and RobustDistiller~\cite{10095480}. The DiT network cross-attends to the learned weighted sum of feature maps of the SSL model, as the features differ in temporal resolutions from the VAE latent.

\vspace{2mm}
\noindent\textbf{Conditioner Network.}
As indicated in Figure~\ref{fig:ditse_architecture}, a feed-forward network using Conformer~\cite{gulati20_interspeech} architecture takes in the VAE latent and transforms it to a cleaner representation that approximates the clean latent. We jointly train the Conditioner network with the main diffusion network, using L1-loss against the ground-truth clean latent alongside the diffusion loss. Our preliminary study shows that stacking this pre-enhanced latent with the diffusion latent can exploit the explicit temporal alignment between the two and thus works more effectively than conditioning via cross-attention. We can view the design as if the diffusion network observes two perspectives of the clean latent at each diffusion step to make its denoising decision -- one corrupted by Gaussian noise, and one corrupted by acoustic conditions translated to the latent space. In addition, we experimented with masking and mapping formulations on the Conditioner branch, where it either outputs a non-negative mask to be multiplied with the input latent or directly outputs the target latent.


\vspace{2mm}
\noindent\textbf{Auxiliary Timestep Embedding.} 
Building on the timestep conditioning described earlier, we additionally concatenate the timestep embedding with the input of the diffusion network. The practice enhances the model's capability to adaptively fuse the diffusion latent and the pre-enhanced latent at different diffusion steps. By intuition, the pre-enhanced latent provides a direct path for structuring a speech envelope in the early diffusion steps, but introduces distractions caused by its own errors and artifacts in the later steps. Without this auxiliary timestep embedding, we noticed from qualitative assessments that residuals of the noise and reverb in the input audio are more likely to be carried over to the final output.

\subsection{Post-hoc Quantization}
While the design above achieves good content accuracy and speaker preservation across audio samples, it exhibits instability in acoustic quality compared to discrete token generation methods. We hypothesize two main causes: (1) a weak correlation between latent space distance and perceptual quality, making small numerical errors perceptually significant; and (2) sensitivity of the diffusion process to noise and sampling, leading to error accumulation over the trajectory for out-of-domain acoustic scenarios. While prior work~\cite{universepp} addressed perceptual issues by fine-tuning the final diffusion steps, we propose to leverage discrete generation methods post-hoc for their benefits in audio fidelity. Specifically, we decode the enhanced VAE latent into audio and use the discrete neural codec DAC~\cite{NEURIPS2023_58d0e78c} model to reconstruct the audio, eliminating audible artifacts and regularizing harmonics of the speech from the latent diffusion generation.

\section{Experiments}
\label{sec:typestyle}

\subsection{Experiment Setup}\label{subsec:expsetup}

\noindent\textbf{Training Datasets.} Our models are trained on publicly available datasets that simulate real-world acoustic conditions. Note that none of these datasets are used for evaluation. For clean speech, we use LibriTTS-R~\cite{koizumi23_interspeech}, a quality-improved 585-hour dataset, further upsampled to \SI{48}{kHz} sample rate via bandwidth extension~\cite{9413575}. Noise samples are uniformly sampled from three datasets: (1) SFS-Static-Dataset~\cite{pmlr-v164-chen22b}, (2) TAU Urban Audio-Visual Scenes 2021 dataset~\cite{9415085}, and (3) the DNS Challenge~\cite{10474162}. Reverberation is simulated using room impulse responses (RIRs) from the OpenSLR28 dataset~\cite{7953152}, MIT IR Survey~\cite{doi:10.1073/pnas.1612524113}, and EchoThief~\cite{echothief}. To generate the pairs of synthetic training data, we employ a data simulation and augmentation procedure based on previous work~\cite{genhancer}. A speech signal is first convolved with a RIR and then mixed with one or two noise samples at a signal-to-noise ratio (SNR) uniformly sampled between -10 dB and 20 dB. The signal also undergoes random equalization and bandwidth limitation.

\noindent\textbf{Hyperparameters and Hardware.} The DiT architecture consists of 12 Transformer layers with 8 attention heads (335M parameters, 1024 channels, conditioned on WavLM features). We also do ablation with a small version (112M parameters, 512 channels, conditioned on RobustDistiller features~\cite{10095480}, distilled from WavLM). Sinusoidal positional encodings are used across all attention layers. The conditioner branch uses a Conformer model of 12 layers, each with an input size, feed-forward dimension, and kernel size of 256, 1024, and 31, respectively. We apply Classifier-Free Guidance~\cite{ho2022classifier} during training with a probability of 10\% to encourage the model to effectively leverage each conditioning feature. We train our models on eight NVIDIA A100 GPUs for 400k steps using a batch size of 128 $\times$ 5 seconds together with the AdamW optimizer. The learning rate is linearly warmed up from 0 to $10^{-4}$ over the first 10\% training steps before decaying to $10^{-5}$ in subsequent steps. We chose the cosine noise scheduler after qualitatively comparing the generated results to ones from other noise schedulers. We adopt $N = 32$ steps at inference as additional steps yield minimal improvements, and use dpmpp-3m-sde sampler from k-diffusion\footnote{https://github.com/crowsonkb/k-diffusion} library.

\begin{table}[tb]
    \centering
    \caption{Objective Metrics on the DAPS dataset for Ablation Studies.}
    \label{tab:ablation}
    \resizebox{\linewidth}{!}{%
        \begin{tabular}{lccccc}
            \toprule
             Experiment & WER $(\downarrow)$ & WB-PESQ $(\uparrow)$ & ESTOI $(\uparrow)$ & DNSMOS $(\uparrow)$ \\
             \midrule
            Input & 5.03 & 1.43  & 66.83 & 2.49 \\
            \midrule
            \multicolumn{6}{c}{Ablation Studies} \\
            \midrule

            Baseline & \makecell{ 4.29 \\ {\scriptsize \textcolor{gray}{$\pm$ 0.71 }}} & \makecell{ 2.31 \\ {\scriptsize \textcolor{gray}{$\pm$ 0.01 }}} & \makecell{ 84.24 \\ {\scriptsize \textcolor{gray}{$\pm$ 0.06 }}} & \makecell{ 3.32 \\ {\scriptsize \textcolor{gray}{$\pm$ 0.01 }}} \\

            \hdashline
            \quad (+) Concat $\alpha_t$ & \makecell{ 4.51 \\ {\scriptsize \textcolor{gray}{$\pm$ 0.8 }}} & \makecell{ 2.42 \\ {\scriptsize \textcolor{gray}{$\pm$ 0.01 }}} & \makecell{ 85.2 \\ {\scriptsize \textcolor{gray}{$\pm$ 0.05 }}} & \makecell{ 3.33 \\ {\scriptsize \textcolor{gray}{$\pm$ 0.01 }}} \\
            \hdashline
            \quad (+) Conditioner {\tiny [Mapping]} &  \makecell{ 3.72 \\ {\scriptsize \textcolor{gray}{$\pm$ 0.39 }}} & \makecell{ 2.43 \\ {\scriptsize \textcolor{gray}{$\pm$ 0.02 }}} & \makecell{ 85.31 \\ {\scriptsize \textcolor{gray}{$\pm$ 0.09 }}} & \makecell{ 3.33 \\ {\scriptsize \textcolor{gray}{$\pm$ 0 }}} \\
    
            \hdashline

            \quad (+) Conditioner {\tiny [Masking]} & \makecell{ 3.62 \\ {\scriptsize \textcolor{gray}{$\pm$ 0.27 }}} & \makecell{ 2.42 \\ {\scriptsize \textcolor{gray}{$\pm$ 0.01 }}} & \makecell{ 85.17 \\ {\scriptsize \textcolor{gray}{$\pm$ 0.07 }}} & \makecell{ 3.34 \\ {\scriptsize \textcolor{gray}{$\pm$ 0.01 }}}\\

            \hdashline
            \quad (+) Post-hoc Quantization & \makecell{ 4.01 \\ {\scriptsize \textcolor{gray}{$\pm$ 0.61 }}} & \makecell{ 2.35 \\ {\scriptsize \textcolor{gray}{$\pm$ 0.01 }}} & \makecell{ 85.15 \\ {\scriptsize \textcolor{gray}{$\pm$ 0.09 }}} & \makecell{ 3.32 \\ {\scriptsize \textcolor{gray}{$\pm$ 0.01 }}} \\

            \midrule
            
            \quad (-) Scaling Down & \makecell{ 8.49 \\ {\scriptsize \textcolor{gray}{$\pm$ 0.45 }}} & \makecell{ 2.12 \\ {\scriptsize \textcolor{gray}{$\pm$ 0.01 }}} & \makecell{ 81.92 \\ {\scriptsize \textcolor{gray}{$\pm$ 0.11 }}} & \makecell{ 3.31 \\ {\scriptsize \textcolor{gray}{$\pm$ 0.01 }}} \\
            
            \midrule[\heavyrulewidth]
            \bottomrule
        \end{tabular}
    }
\end{table}

\begin{table*}[t]
    \centering
    \caption{Comparison of our methods and state-of-the-art models on the three evaluation axes. \textbf{Bold} numbers highlights the best results, and \underline{underline} numbers the second best. Entries are skipped for evaluation sets that do not provide ground-truth clean audio.}
    \label{tab:listening_tests}
    \resizebox{\linewidth}{!}{%
        \begin{tabular}{lcccccccccc}
            \toprule
            && \multicolumn{3}{c}{Content [WER~$\downarrow$]} & \multicolumn{2}{c}{Speaker [MOS~$\uparrow$]} & \multicolumn{4}{c}{Quality [MOS~$\uparrow$]}\\
            \cmidrule(lr){3-5}
            \cmidrule(lr){6-7}
            \cmidrule(lr){8-11}
             & \makecell{SR  \\ {\scriptsize (kHz)}} & \makecell{DAPS  \\ {\scriptsize (16 kHz)}} & \makecell{VBD  \\ {\scriptsize (16 kHz)}} & \makecell{EARS  \\ {\scriptsize (16 kHz)}} & \makecell{VBD  \\ {\scriptsize (48 kHz)}} & \makecell{EARS  \\ {\scriptsize (48 kHz)}} & \makecell{DAPS  \\ {\scriptsize (48 kHz)}} & \makecell{AQECC  \\ {\scriptsize (16 kHz)}} & \makecell{DEMO  \\ {\scriptsize (16 kHz)}} & \makecell{VBD  \\ {\scriptsize (48 kHz)}} \\
            \midrule
            Input & 48 & 5.03 & 5.70 & 15.08 & 4.84 {\scriptsize \textcolor{gray}{$\pm$ 0.02}} & 4.89 {\scriptsize \textcolor{gray}{$\pm$ 0.02}} & 1.74 {\scriptsize \textcolor{gray}{$\pm$ 0.04}} & 2.74 {\scriptsize \textcolor{gray}{$\pm$ 0.06}} & 1.93 {\scriptsize \textcolor{gray}{$\pm$ 0.05}} & 1.84 {\scriptsize \textcolor{gray}{$\pm$ 0.05}} \\
            Clean & 48 & --- & --- & --- & 4.41 {\scriptsize \textcolor{gray}{$\pm$ 0.04}}  & 4.18 {\scriptsize \textcolor{gray}{$\pm$ 0.05}}  & 4.30 {\scriptsize \textcolor{gray}{$\pm$ 0.04}} & --- & --- & 3.87 {\scriptsize \textcolor{gray}{$\pm$ 0.05}} \\
            VAE Reconst. & 48 & 0.87 & 1.23 & 3.34 & 4.28 {\scriptsize \textcolor{gray}{$\pm$ 0.04}}  & 4.10 {\scriptsize \textcolor{gray}{$\pm$ 0.05}} & 4.25 {\scriptsize \textcolor{gray}{$\pm$ 0.04}} & --- & --- & 3.54 {\scriptsize \textcolor{gray}{$\pm$ 0.05}} \\

            \midrule
            HiFi-GAN-2~\cite{su2021hifi} & 48 & 9.16 & 6.63 & \textbf{23.67} & 4.09 {\scriptsize \textcolor{gray}{$\pm$ 0.05}} & 3.16 {\scriptsize \textcolor{gray}{$\pm$ 0.07}} & 3.63 {\scriptsize \textcolor{gray}{$\pm$ 0.05}} & 3.70 {\scriptsize \textcolor{gray}{$\pm$ 0.05}} & 3.33 {\scriptsize \textcolor{gray}{$\pm$ 0.06}} & 3.77 {\scriptsize \textcolor{gray}{$\pm$ 0.05}}\\
            Miipher~\cite{koizumi2023miipher} & 22.1 & 7.64 & 12.02 & 32.32 & 3.24 {\scriptsize \textcolor{gray}{$\pm$ 0.06}} & 2.68 {\scriptsize \textcolor{gray}{$\pm$ 0.06}} & 3.40 {\scriptsize \textcolor{gray}{$\pm$ 0.05}} & 3.53 {\scriptsize \textcolor{gray}{$\pm$ 0.05}} & 3.15 {\scriptsize \textcolor{gray}{$\pm$ 0.05}} & 3.42 {\scriptsize \textcolor{gray}{$\pm$ 0.06}} \\
            Genhancer~\cite{genhancer} & 44.1 & 6.04 & 6.47 & \underline{26.33} & 4.03 {\scriptsize \textcolor{gray}{$\pm$ 0.05}} & 3.27 {\scriptsize \textcolor{gray}{$\pm$ 0.07}} & 4.08 {\scriptsize \textcolor{gray}{$\pm$ 0.05}} & \underline{3.89} {\scriptsize \textcolor{gray}{$\pm$ 0.04}} & \textbf{3.78} {\scriptsize \textcolor{gray}{$\pm$ 0.05}} & \textbf{4.03} {\scriptsize \textcolor{gray}{$\pm$ 0.05}} \\
            \midrule
            SGMSE+~\cite{richter2023speech} & 48 & 6.69 & 10.32 & 33.60 & 3.87 {\scriptsize \textcolor{gray}{$\pm$ 0.06}} & 3.03 {\scriptsize \textcolor{gray}{$\pm$ 0.06}} & 3.51 {\scriptsize \textcolor{gray}{$\pm$ 0.05}} & 3.71 {\scriptsize \textcolor{gray}{$\pm$ 0.04}} & 3.33 {\scriptsize \textcolor{gray}{$\pm$ 0.05}} & 3.59 {\scriptsize \textcolor{gray}{$\pm$ 0.06}} \\
            StoRM~\cite{lemercier2023storm} & 16 & 11.27 & 9.86 & 45.87 & 3.54 {\scriptsize \textcolor{gray}{$\pm$ 0.06}} & 2.75 {\scriptsize \textcolor{gray}{$\pm$ 0.06}}  & 2.62 {\scriptsize \textcolor{gray}{$\pm$ 0.05}} & 3.61 {\scriptsize \textcolor{gray}{$\pm$ 0.05}} & 2.53 {\scriptsize \textcolor{gray}{$\pm$ 0.06}} & 2.94 {\scriptsize \textcolor{gray}{$\pm$ 0.05}} \\

            \midrule

            (\textbf{Ours}) DiTSE & 48 & \textbf{3.56} & \textbf{4.93} & 26.42 & \underline{4.20} {\scriptsize \textcolor{gray}{$\pm$ 0.04}} & \underline{3.45} {\scriptsize \textcolor{gray}{$\pm$ 0.06}} & \textbf{4.34} {\scriptsize \textcolor{gray}{$\pm$ 0.04}} & 3.88 {\scriptsize \textcolor{gray}{$\pm$ 0.05}} & 3.61 {\scriptsize \textcolor{gray}{$\pm$ 0.05}} & 3.91 {\scriptsize \textcolor{gray}{$\pm$ 0.05}} \\
            (\textbf{Ours}) DiTSE +  Post & 48 & \underline{3.71} & \underline{5.39} & 26.72 & \textbf{4.27} {\scriptsize \textcolor{gray}{$\pm$ 0.04}} & \textbf{3.50} {\scriptsize \textcolor{gray}{$\pm$ 0.06}} & \underline{4.32} {\scriptsize \textcolor{gray}{$\pm$ 0.04}} & \textbf{3.97} {\scriptsize \textcolor{gray}{$\pm$ 0.04}} & \underline{3.77} {\scriptsize \textcolor{gray}{$\pm$ 0.05}} & \underline{4.00} {\scriptsize \textcolor{gray}{$\pm$ 0.05}}\\

            \midrule[\heavyrulewidth]
            \bottomrule
        \end{tabular} 
    }
\end{table*}

\subsection{Ablation Study}
We investigate the design choices of the proposed DiTSE model on the DAPS dataset~\cite{6981922}, which consists of pairs of studio-quality speech recordings and their degraded versions under real-world acoustic conditions. Our analysis utilizes a subset of DAPS as used by prior work~\cite{genhancer} -- one female voice (\textit{f10}) and one male voice (\textit{m10}) reading a 2-minute script (\textit{script5}).
Table~\ref{tab:ablation} presents the objective scores of Word Error Rate (WER)\footnote{Calculated between the enhanced audio and the clean reference audio transcribed by Whisper-Large-v3-v20240930 (\url{https://github.com/openai/whisper})}, WideBand Perceptual Evaluation of Speech Quality (WB-PESQ), Extended Short-Time Objective Intelligibility (ESTOI), and the non-intrusive metric DNSMOS~\cite{9746108}. Given the stochastic nature of generative approaches, we perform inference five times for each model using different random seeds and report the statistics across all the runs.

Our \textbf{Baseline} approach, a DiT model without auxiliary timestep embeddings or Conditioner network, achieved notable content preservation.
Incorporating additional timestep embeddings at the input improves acoustic quality.
Introducing Conditioner network further improves WER, and the masking formulation shows better stability than the mapping formulation in the latent space. Lastly, applying the post-hoc quantization to the enhanced audio from our full-sized model improves its perceived quality, but compromises content fidelity slightly.

While our model is larger than the baselines, the comparison remains valid and highlights the strengths of our approach. We also evaluated a 4.5 times smaller version, which showed reduced content accuracy but maintained overall quality comparable to diffusion-based baselines. Although still about 2 times larger than SGMSE+ in terms of parameters, our model benefits from a DiT-based architecture that diffuses in the VAE’s latent space, allowing more efficient training (e.g., four times larger batch sizes) and significantly better acoustic quality with scaling, an advantage that is less evident in U-Net-based generative models on time-frequency representations.

\subsection{Three-axis Evaluations}
We compare the proposed methods to state-of-the-art baselines using objective and subjective evaluations. 
We consider three evaluation axes: (1) content accuracy measured by WER; (2) acoustic quality measured by quality mean-opinion-score (MOS); (3) speaker preservation measured by similarity MOS. We examine the methods on five evaluation sets unseen during training: DAPS Real, AQECC, and DEMO, as utilized in prior work~\cite{genhancer}, each consisting of 100 samples representing diverse real-world scenarios; VoiceBank-DEMAND (VBD)~\cite{valentinibotinhao16_ssw} for denoising benchmark; and EARS~\cite{richter2024ears} for challenging denoising and de-reverberation scenarios. For VBD and EARS, we sample twenty distinct speakers in each category with two samples per speaker, ensuring a variety of voices with balanced genders. We compare with the following methods: HiFi-GAN-2~\cite{su2021hifi}; transcription-free Miipher~\cite{koizumi2023miipher}; Genhancer~\cite{genhancer}, a discrete token generation method that is closest to our method in conditioning designs; SGMSE+~\cite{richter2023speech} and StoRM~\cite{lemercier2023storm} as diffusion baselines. For consistency, all baseline models are trained using the same dataset as our model (section \ref{subsec:expsetup}).

We conducted MOS listening tests via Prolific\footnote{https://www.prolific.co/.} to evaluate audio quality and speaker similarity. For quality, listeners rated samples from 1 (Bad) to 5 (Excellent) for cleanness and naturalness, using a fixed clean DAPS sample as the reference of studio audio quality. For speaker similarity, listeners rated the preservation of the speaker's voice in a sample with respect to the corresponding input, using the same scale. Each method received at least 400 ratings for each evaluation set.

We use the highest available sampling rates for each model and dataset. We also include \SI{16}{kHz} and \SI{22}{kHz} downsampled versions of our results for fair comparison to confirm that performance differences are not driven by sampling rates. Detailed results per sampling rate are available on our demo page.

Table~\ref{tab:listening_tests} reports the results for the three evaluation axes. 
DiTSE stands out as the only model capable of reducing content error compared to the noisy input on DAPS and VBD, significantly outperforming all the baselines. 
EARS is a challenging dataset with severe noise and reverberation and barely intelligible speech. It also includes complex out-of-domain sound, such as fillers, emotional interjections and laughter. DiTSE ranks among the top generative methods on EARS, while the regression-based HiFi-GAN-2, as it modifies the original waveform, yields slightly higher content accuracy but noticeably lower perceptual quality.
Striking the right balance between enabling content restoration and refraining models from hallucinations still remains an open problem under extreme SNR scenarios.

Our methods significantly improve acoustic quality over all the baselines.
To the best of our knowledge, \textbf{DiTSE-\textit{Base}} is the first method that achieves a Quality MOS of 4.34 that is indistinguishable from the real studio-quality audio of DAPS.
It achieves an MOS comparable with Genhancer on AQECC, but under-performs on DEMO and VBD. We pinpoint that the VAE could limit audio quality as the reconstruction of clean audio (\textbf{VAE Reconst.}) displays notable reduction in Quality MOS on VBD. 
Incorporating the post-hoc quantization (\textbf{DiTSE + Post}) improves audio fidelity by regularizing harmonics of the generated speech and eliminating generation artifacts, achieving state-of-the-art audio quality across all the evaluation sets. The content remains largely unchanged aside from a few subtle pronunciation variations at difficult spots.

Lastly, Speaker MOS shows significant advantages of DiTSE in preserving speaker and paralinguistic features over baselines.
The robust conditioning enables the model to effectively attend to speaker attributes in complex acoustic scenarios. Furthermore, generation in the continuous space facilitates the preservation of spoken nuances such as breath and coarticulation, resulting in more natural speech compared to generative methods using discrete tokens, while quantization after continuous-space generation enhances voice fidelity.

\section{Conclusions}
\label{sec:conclusion}

We introduce DiTSE, a speech enhancement approach built upon a latent diffusion transformer model. DiTSE effectively preserves spoken content and speaker characteristics by incorporating robust conditioning features. We additionally leverage post-hoc quantization to mitigate audio fidelity issues of diffusion-based approaches. Comprehensive evaluations demonstrate that our proposed approach efficiently achieves state-of-the-art audio quality, for the first time comparable to real studio quality, in 32 generation steps, while significantly improving content accuracy and speaker consistency over previous methods. Diffusion distillation techniques~\cite{Yin_2024_CVPR} could help reduce the computational cost of DiTSE in the future.

\bibliographystyle{IEEEbib}
\bibliography{refs}

\begin{thebibliography}{10}

\bibitem{o2024speech}
Douglas O'Shaughnessy,
\newblock ``Speech enhancement—a review of modern methods,''
\newblock {\em IEEE Transactions on Human-Machine Systems}, 2024.

\bibitem{wang2023selm}
Ziqian Wang, Xinfa Zhu, Zihan Zhang, et~al.,
\newblock ``{SELM}: Speech enhancement using discrete tokens and language models,''
\newblock {\em arXiv preprint arXiv:2312.09747}, 2023.

\bibitem{xue2023low}
Huaying Xue, Xiulian Peng, and Yan Lu,
\newblock ``Low-latency speech enhancement via speech token generation,''
\newblock {\em arXiv preprint arXiv:2310.08981}, 2023.

\bibitem{genhancer}
Haici Yang, Jiaqi Su, Minje Kim, et~al.,
\newblock ``Genhancer: High-fidelity speech enhancement via generative modeling on discrete codec tokens,''
\newblock in {\em Interspeech}, 2024.

\bibitem{richter2023speech}
Julius Richter, Simon Welker, Jean-Marie Lemercier, et~al.,
\newblock ``Speech enhancement and dereverberation with diffusion-based generative models,''
\newblock {\em IEEE/ACM Transactions on Audio, Speech, and Language Processing}, vol. 31, 2023.

\bibitem{lu2022conditional}
Yen-Ju Lu, Zhong-Qiu Wang, Shinji Watanabe, et~al.,
\newblock ``Conditional diffusion probabilistic model for speech enhancement,''
\newblock in {\em ICASSP}. IEEE, 2022.

\bibitem{serra2022universal}
Joan Serr{\`a}, Santiago Pascual, Jordi Pons, et~al.,
\newblock ``Universal speech enhancement with score-based diffusion,''
\newblock {\em arXiv preprint arXiv:2206.03065}, 2022.

\bibitem{universepp}
Robin Scheibler, Yusuke Fujita, Yuma Shirahata, et~al.,
\newblock ``Universal score-based speech enhancement with high content preservation,''
\newblock in {\em Interspeech}, 2024, pp. 1165--1169.

\bibitem{lemercier2023storm}
Jean-Marie Lemercier, Julius Richter, Simon Welker, et~al.,
\newblock ``Storm: A diffusion-based stochastic regeneration model for speech enhancement and dereverberation,''
\newblock {\em IEEE/ACM Transactions on Audio, Speech, and Language Processing}, 2023.

\bibitem{10447017}
Muqiao Yang, Chunlei Zhang, Yong Xu, et~al.,
\newblock ``usee: Unified speech enhancement and editing with conditional diffusion models,''
\newblock in {\em ICASSP}, 2024.

\bibitem{dhyani2025high}
Tushar Dhyani, Florian Lux, Michele Mancusi, and Fothers,
\newblock ``High-resolution speech restoration with latent diffusion model,''
\newblock in {\em ICASSP}. IEEE, 2025.

\bibitem{kumar-etal-2025-prose}
Sonal Kumar, Sreyan Ghosh, Utkarsh Tyagi, et~al.,
\newblock ``{P}ro{SE}: Diffusion priors for speech enhancement,''
\newblock in {\em NAACL}, 2025, pp. 12470--12483.

\bibitem{6981922}
Gautham~J. Mysore,
\newblock ``Can we automatically transform speech recorded on common consumer devices in real-world environments into professional production quality speech?—a dataset, insights, and challenges,''
\newblock {\em IEEE SPL}, 2015.

\bibitem{evans2024stable}
Zach Evans, Julian~D Parker, CJ~Carr, et~al.,
\newblock ``Stable audio open,''
\newblock {\em arXiv preprint arXiv:2407.14358}, 2024.

\bibitem{hoogeboom2023simple}
Emiel Hoogeboom, Jonathan Heek, and Tim Salimans,
\newblock ``simple diffusion: End-to-end diffusion for high resolution images,''
\newblock in {\em ICML}. PMLR, 2023.

\bibitem{salimans2022progressive}
Tim Salimans and Jonathan Ho,
\newblock ``Progressive distillation for fast sampling of diffusion models,''
\newblock in {\em ICLR}, 2022.

\bibitem{NEURIPS2023_58d0e78c}
Rithesh Kumar, Prem Seetharaman, Alejandro Luebs, Ishaan Kumar, and Kundan Kumar,
\newblock ``High-fidelity audio compression with improved rvqgan,''
\newblock in {\em NeurIPS}, 2023, vol.~36.

\bibitem{welker2022speech}
Simon Welker, Julius Richter, and Timo Gerkmann,
\newblock ``Speech enhancement with score-based generative models in the complex stft domain,''
\newblock in {\em Interspeech}, 2022.

\bibitem{lovelace24_interspeech}
Justin Lovelace, Soham Ray, Kwangyoun Kim, et~al.,
\newblock ``Sample-efficient diffusion for text-to-speech synthesis,''
\newblock in {\em Interspeech}, 2024.

\bibitem{peebles2023scalable}
William Peebles and Saining Xie,
\newblock ``Scalable diffusion models with transformers,''
\newblock in {\em IEEE/CVF ICCV}, 2023.

\bibitem{9814838}
Sanyuan Chen, Chengyi Wang, Zhengyang Chen, et~al.,
\newblock ``Wavlm: Large-scale self-supervised pre-training for full stack speech processing,''
\newblock {\em IEEE Journal of Selected Topics in Signal Processing}, vol. 16, no. 6, 2022.

\bibitem{10095480}
H.~R. Guimarães, A.~Pimentel, A.~R. Avila, et~al.,
\newblock ``Robustdistiller: Compressing universal speech representations for enhanced environment robustness,''
\newblock in {\em ICASSP}, 2023.

\bibitem{gulati20_interspeech}
Anmol Gulati, James Qin, Chung-Cheng Chiu, et~al.,
\newblock ``Conformer: Convolution-augmented transformer for speech recognition,''
\newblock in {\em Interspeech}, 2020.

\bibitem{koizumi23_interspeech}
Yuma Koizumi, Heiga Zen, Shigeki Karita, et~al.,
\newblock ``Libritts-r: A restored multi-speaker text-to-speech corpus,''
\newblock in {\em Interspeech}, 2023, pp. 5496--5500.

\bibitem{9413575}
Jiaqi Su, Yunyun Wang, Adam Finkelstein, et~al.,
\newblock ``Bandwidth extension is all you need,''
\newblock in {\em ICASSP}, 2021.

\bibitem{pmlr-v164-chen22b}
Ziyang Chen, Xixi Hu, and Andrew Owens,
\newblock ``Structure from silence: Learning scene structure from ambient sound,''
\newblock in {\em 5th Conference on Robot Learning}. 2022, PMLR.

\bibitem{9415085}
Shanshan Wang, Annamaria Mesaros, Toni Heittola, et~al.,
\newblock ``A curated dataset of urban scenes for audio-visual scene analysis,''
\newblock in {\em ICASSP}, 2021.

\bibitem{10474162}
H.~Dubey, A.~Aazami, V.~Gopal, et~al.,
\newblock ``Icassp 2023 deep noise suppression challenge,''
\newblock {\em IEEE OJSP}, 2024.

\bibitem{7953152}
Tom Ko, Vijayaditya Peddinti, Daniel Povey, et~al.,
\newblock ``A study on data augmentation of reverberant speech for robust speech recognition,''
\newblock in {\em ICASSP}, 2017.

\bibitem{doi:10.1073/pnas.1612524113}
James Traer and Josh~H. McDermott,
\newblock ``Statistics of natural reverberation enable perceptual separation of sound and space,''
\newblock {\em Proceedings of the National Academy of Sciences}, 2016.

\bibitem{echothief}
``{EchoThief [dataset]},'' \url{https://www.echothief.com/echothief/},
\newblock [Accessed: 2025-09-17].

\bibitem{ho2022classifier}
Jonathan Ho and Tim Salimans,
\newblock ``Classifier-free diffusion guidance,''
\newblock {\em arXiv preprint arXiv:2207.12598}, 2022.

\bibitem{su2021hifi}
Jiaqi Su, Zeyu Jin, and Adam Finkelstein,
\newblock ``Hifi-gan-2: Studio-quality speech enhancement via generative adversarial networks conditioned on acoustic features,''
\newblock in {\em IEEE WASPAA}, 2021.

\bibitem{koizumi2023miipher}
Yuma Koizumi, Heiga Zen, Shigeki Karita, et~al.,
\newblock ``Miipher: A robust speech restoration model integrating self-supervised speech and text representations,''
\newblock in {\em IEEE WASPAA}, 2023.

\bibitem{9746108}
Chandan K~A Reddy, Vishak Gopal, and Ross Cutler,
\newblock ``Dnsmos p.835: A non-intrusive perceptual objective speech quality metric to evaluate noise suppressors,''
\newblock in {\em ICASSP}, 2022.

\bibitem{valentinibotinhao16_ssw}
Cassia Valentini-Botinhao, Xin Wang, Shinji Takaki, et~al.,
\newblock ``Investigating rnn-based speech enhancement methods for noise-robust text-to-speech,''
\newblock in {\em ISCA Workshop on Speech Synthesis Workshop}, 2016.

\bibitem{richter2024ears}
Julius Richter, Yi-Chiao Wu, Steven Krenn, et~al.,
\newblock ``{EARS}: An anechoic fullband speech dataset benchmarked for speech enhancement and dereverberation,''
\newblock in {\em Interspeech}, 2024.

\bibitem{Yin_2024_CVPR}
Tianwei Yin, Micha\"el Gharbi, Richard Zhang, et~al.,
\newblock ``One-step diffusion with distribution matching distillation,''
\newblock in {\em IEEE/CVF CVPR}, June 2024, pp. 6613--6623.

\end{thebibliography}

\end{document}